\documentclass[apjl]{emulateapj}
\usepackage{amssymb,graphicx}
\usepackage{epsfig}
\usepackage{psfrag}
\usepackage[usenames]{color}
\usepackage{multirow}
\usepackage{rotating}
\usepackage{hyperref}
\usepackage{ragged2e}

\usepackage{subfigure}
\usepackage{ulem}

\newcommand{\eqref}[1]{(\ref{#1})}
\newcommand{\vp}[1]{{\color{blue} {VASILIS SAYS: #1}}}
\newcommand{\jb}[1]{{\color{red} {JANE SAYS: #1}}}

\newcommand{\comment}[1]{}

\shorttitle{Simulations of accreting, spinning binary black holes in full GR}
\shortauthors{Paschalidis, Bright, Ruiz, \& Gold}

\begin{document}


\title{Minidisk dynamics in accreting, spinning black hole binaries:
  Simulations in full general relativity}

\author{Vasileios Paschalidis${}^{1,2}$, Jane Bright${}^{1}$, Milton
  Ruiz${}^3$, Roman Gold${}^4$}
\affil{
  ${}^1$Department of Astronomy, University of Arizona, Tucson, AZ 85726, USA\\
  ${}^2$Department of Physics, University of Arizona, Tucson, AZ 85726, USA\\
${}^3$Department of Physics, University of Illinois at Urbana-Champaign, Urbana, IL 61801\\
${}^4$CP3 origins $|$ Southern Denmark University (SDU) Campusvej 55, Odense, Denmark\\}


\begin{abstract}

We perform magnetohydrodynamic simulations of accreting, equal-mass
binary black holes in full general relativity focusing on the impact
of black hole spin on the dynamical formation and evolution of
minidisks.
We find that during the late inspiral the sizes of minidisks are
primarily determined by the interplay between the tidal field and the
effective innermost stable orbit around each black hole. Our
calculations support that a minidisk forms when the Hill sphere around
each black hole is significantly larger than the black hole's
effective innermost stable orbit. As the binary inspirals, the radius
of the Hill sphere decreases, and minidisks consequently shrink in
size. As a result, electromagnetic signatures associated with
minidisks may be expected to gradually disappear prior to merger when
there are no more stable orbits within the Hill sphere. In particular,
a gradual disappearance of a hard electromagnetic component in the
spectrum of such systems could provide a characteristic signature of
merging black hole binaries. For a binary of given total mass, the
timescale to minidisk ``evaporation'' should therefore depend on the
black hole spins and the mass ratio. We also demonstrate that
accreting binary black holes with spin have a higher efficiency for
converting accretion power to jet luminosity. These results could
provide new ways to estimate black hole spins in the future.

\end{abstract}


\maketitle

\section{Introduction}

Supermassive black hole (SMBH) binaries are expected to form in gas
rich environments as a result of galaxy mergers
~\citep{Rodriguez:2009}. These systems form unique sources for
multimessenger signals. SMBH binary evolution is particularly
promising as potentially strong sources of gravitational and various
forms of electromagnetic emission. Gravitational waves from SMBH
binaries are expected to be detectable by Pulsar Timing Arrays or by
future space-based gravitational wave observatories such as the Laser
Interferometer Space Antenna (LISA)~\citep{Hobbs:2010,LISA}. Recently,
observational efforts searching for such systems have yielded an
unprecedented number of exciting candidate sources based on various
diagnostics including periodicities in light curves, spectral features
such as double-horned emission lines, and (lensing) flares (see,
e.g.,~\cite{Drake:2009,Graham:2015tba,Charisi:2016fqw,Hu:2019nhk}). A
main challenge in these efforts is our lack of understanding of the
unique features the electromagnetic emission from these systems
has. As a result, much of the theoretical work on accreting binary
black holes to-date has focused on identifying interesting
characteristic electromagnetic signatures that accompany their
gravitational wave signals.

In this paper we investigate the conditions under which minidisks form
around each individual black hole in the late stages of the binary
black hole inspiral. It has been argued that minidisks are responsible
for the emission of the hardest part of the electromagnetic spectrum
due to shock heating~\citep{Sesana:2011zv,Roedig:2014,Farris:2015}.
Emission from the hot minidisks could make up for any ``notch" in the
high frequency spectrum that might be expected due to the lack of gas in
the disk cavity during the inspiral phase. Minidisks could also be
responsible for providing modifications to broad emission-line
profiles~\citep{Nguyen:2019}. The potentially observable
electromagnetic (EM) emission originating from
minidisks~\citep{d'Ascoli:2018} as well as the possibility to use them
as ``smoking-gun'' features to identify SMBH binaries, makes
understanding the conditions under which minidisks form and maintain a
persistent structure toward the late stages of the inspiral a vital
goal. Minidisks have been seen in some recent Newtonian studies,
e.g.,~\cite{Farris:2014, Farris:2015}, and studies with approximate
spacetime metrics~\citep{Bowen:2017, Bowen:2018}, but not in previous
studies in full general relativity, e.g.,~\cite{Farris:2012,
  Gold:2014a,Gold:2014b}), highlighting the need to investigate the
conditions for minidisk formation in relativistic gravitation. The
studies using approximate spacetime metrics focusing on near
equal-mass systems~\citep{Bowen:2017, Bowen:2018, Bowen:2019},
initialize the simulations with minidisks in addition to the
circumbinary disk, but do not report persistent, steady-state
minidisks. Instead, those works report cycles of depletion and
replenishment as well as sloshing of mass between the two minidisks.

Previous theoretical work on circumbinary disks incorporating varying
degrees of relativistic gravity and magnetic fields include
\citep{Giacomazzo:2012, Noble:2012, Farris:2012, Farris:2014,
  Gold:2014a,Gold:2014b, Kelly:2017, Bowen:2018,Khan:2018ejm,
  Bowen:2019,Armengol:2021shd}, but only a subset of these focus on
circumbinary accretion, see~\citep{Gold:2019nqg} for a recent
review. \cite{Gold:2014b} proposed that circumbinary disks form
minidisks whenever there are stable circular orbits within the Hill
sphere around each black hole. In this work we perform simulations in
full general relativity (GR) to test this hypothesis. Our approach
makes no approximation for the spacetime metric and as a result we do
not have to excise parts of the domain and/or impose ad hoc inflow
boundary conditions. Instead, the black holes are resolved objects on
our computational grid. The duration of our simulations are
significantly shorter than in some earlier treatments. However, the
regime of minidisks that is probed here exhibits much shorter
relaxation times.

The analysis we follow is guided in part by the restricted three-body
Newtonian problem, where one can define regions in which gravity is
dominated by each of the binary components, as quantified by the Hill
sphere $r_{\rm Hill}=0.5 (q/3)^{1/3} d$. Here $q$ is the mass ratio
and $d$ is the binary separation. While the systems studied here are
not in the Newtonian regime, it is still expected that many Newtonian
aspects carry over to the relativistic regime with some appropriate
corrections. On the other hand, it is well-known that stable circular
orbits (and hence disks) around general relativistic black holes can
exist only outside the innermost stable circular orbit (ISCO) $r_{\rm
  ISCO}=r_{\rm ISCO}(\chi)$ where $\chi$ is the dimensionless black
hole spin parameter satisfying $|\chi|\leq 1$. Therefore, the
expectation is that minidisks can form early on in the inspiral at
binary separations that satisfy $r_{\rm Hill} \gg r_{\rm ISCO}(\chi)$,
so that stable orbits around each black hole can exist within the Hill
sphere~\citep{Gold:2014b}. However, in the later stages of the
inspiral when $r_{\rm Hill} \sim r_{\rm ISCO}(\chi)$ no stable orbits
exist within the Hill sphere and matter is quickly accreted as soon as
it enters the Hill sphere.  For an equal-mass, nonspinning system, we
can perform a simple Newtonian estimate which yields $r_{\rm Hill} =
0.5 (q/3)^{1/3} d \sim 0.35 d$ and $r_{\rm ISCO} = 3 M$, setting these
equal gives a threshold separation for minidisks to exist $d_{\rm
  thres}\simeq 8.7 M$.  In particular, one would require $d\gg d_{\rm
  thres}$ for persistent minidisks to exist. As the binary inspirals
$d$ approaches $d_{\rm thres}$, and any persistent minidisks should
evaporate.

Astrophysical black holes are expected to have spin~\citep{Lynden-Bell:1969}.
X-ray reflection measurement techniques have been used in over two dozen AGN
systems to determine the spins of the SMBHs, and have found that the majority
of these systems are rapidly rotating with $\chi \ge 0.9$, while the most
massive SMBHs ($\gtrsim 10^8M_{\odot}$) have slower, but still significant
spins in the range of $\chi \sim 0.5 - 0.7$~\citep{Reynolds:2019}. As spin is
crucial to determine the ISCO, incorporating spin into our study of minidisks
is important both for astrophysical implications, e.g., jets, and for providing
a clear diagnostic for testing our hypothesis regarding minidisk formation.

By performing magnetohydrodynamic simulations in full GR of $q=1$ binary
black holes with spins both (anti)aligned
with the orbital angular momentum, and one aligned and the other
antialigned we demonstrate explicitly the impact of spin in allowing
the formation of minidisks, thereby confirming the expectations
from~\cite{Gold:2014b} that minidisks should form only when stable
orbits exist within the Hill sphere around each black hole. In
addition, we demonstrate the importance of spin on jet outflows as we
find that spinning binaries have stronger jet luminosities.

In the fully relativistic circumbinary accretion studies of
\cite{Farris:2012,Gold:2014a,Gold:2014b} the launching of jets from
the interaction of magnetic fields with black hole horizons, even for
nonspinning binary black holes, had already highlighted the importance
of full GR in discovering features that are entirely missed in
Newtonian studies, and studies that include varying degrees of
relativity, but still do not capture black hole horizons. Minidisk
evaporation as discussed in this work highlights yet another effect in
which GR plays a key role in determining even the qualitative
behavior.

The rest of the paper is structured as follows. In
Section~\ref{methods} we present the initial data, grids, and
numerical methods we adopt for our evolutions. In
Section~\ref{results_and_discussion} we present our results, and
summarize in Section~\ref{conclusions}. We denote the total binary
mass $M$ and both the primary and secondary mass $m$. We set $G=c=1$.

\section{Methods and initial data}
\label{methods}

In this work we evolve the spacetime metric by solving Einstein's
field equations in the BSSN formulation
\citep{Baumgarte:1998te,Shibata:1995}. Our methods for solving
Einstein's field equations for the gravitational field, and the
equations of ideal magnetohydrodynamics in curved spacetime have been
described previously in \citep{Khan:2018ejm}, where we refer the
reader for more details and further references. Our code has been
extensively tested against the majority of GR magnetohydrodynamics
(GRMHD) codes in the context of single black hole spacetimes in
\citep{Porth:2019}. As in~\citep{Khan:2018ejm} we use puncture initial
data for the spacetime metric adopting the {\tt TwoPunctures}
code~\citep{Ansorg:2004ds}, but set the black holes initially on a
quasicircular orbit at a larger coordinate separation of 20M. We also
treat black hole spin for the first time in full GR calculations of
circumbinary accretion. We consider equal-mass binaries in four spin
configurations: $\chi_1=\chi_2=0$ (nonspinning case labeled
$\chi_{00}$), $\chi_1=\chi_2=0.75$ (case $\chi_{++}$),
$\chi_1=\chi_2=-0.75$ (case $\chi_{--}$), $\chi_1=-\chi_2=0.75$ (case
$\chi_{+-}$), where $\chi_1$, and $\chi_2$ are the dimensionless spins
of each black hole, and the $+$ ($-$) sign indicates spin aligned
(antialigned) with the orbital angular momentum. The matter and
magnetic field initial data are identical to those
in~\citep{Gold:2014a} and the same for all binary configurations we
study. The circumbinary disk and binary angular momenta are aligned.
The equation of state corresponds to a $\Gamma=4/3$ $\Gamma$-law, and
there is no cooling employed during the evolution. We employ three sets of
nested levels of mesh refinement: one centered on each black hole and
one centered at the origin. The set centered on the origin has 10
levels of refinement, while those centered on the black holes have 12
levels for case $\chi_{00}$, and 13 levels for cases with nonzero
spin. We place the outer boundary at 3072$M$, and the resolution on
the coarsest refinement level to 48$M$. The finest level resolution is
$\simeq M/85$ ($\simeq M/43$) for cases with (without) spin. The black
hole apparent horizon diameters are resolved by $\gtrsim 51$ ($\gtrsim
41$) grid points in their smallest dimension for the spinning
(nonspinning) cases. The half-side lengths of the refinement boxes are
given by $3072/2^n,\ n=0,1,\ldots 12$, where $n$ indexes the levels of
refinement with $n=0$ denoting the coarsest level.

\section{Results }
\label{results_and_discussion}

Nonspinning black holes have an ISCO (areal) radius of $r_{\rm ISCO}
= 6m=3M$.  Prograde (retrograde) orbits around black holes with spin
$\chi=0.75$ have smaller (larger) $r_{\rm ISCO} = 3.16m= 1.58M$
($r_{\rm ISCO} = 8.28m= 4.14M$), thereby allowing more (less) space
for stable orbits within the black hole Hill sphere, making it easier
(more difficult) for minidisks to form. We use these radii as
coordinate radii in our figures to indicate the location of the ISCO
around each black hole. We point out that this is neither a gauge
invariant, nor is it a precise measure for relativistic binaries. We
only use these radii as an approximate way to visualize the ISCO in
our figures.

\begin{figure*}
  \centering

  \includegraphics[width=0.99\textwidth]{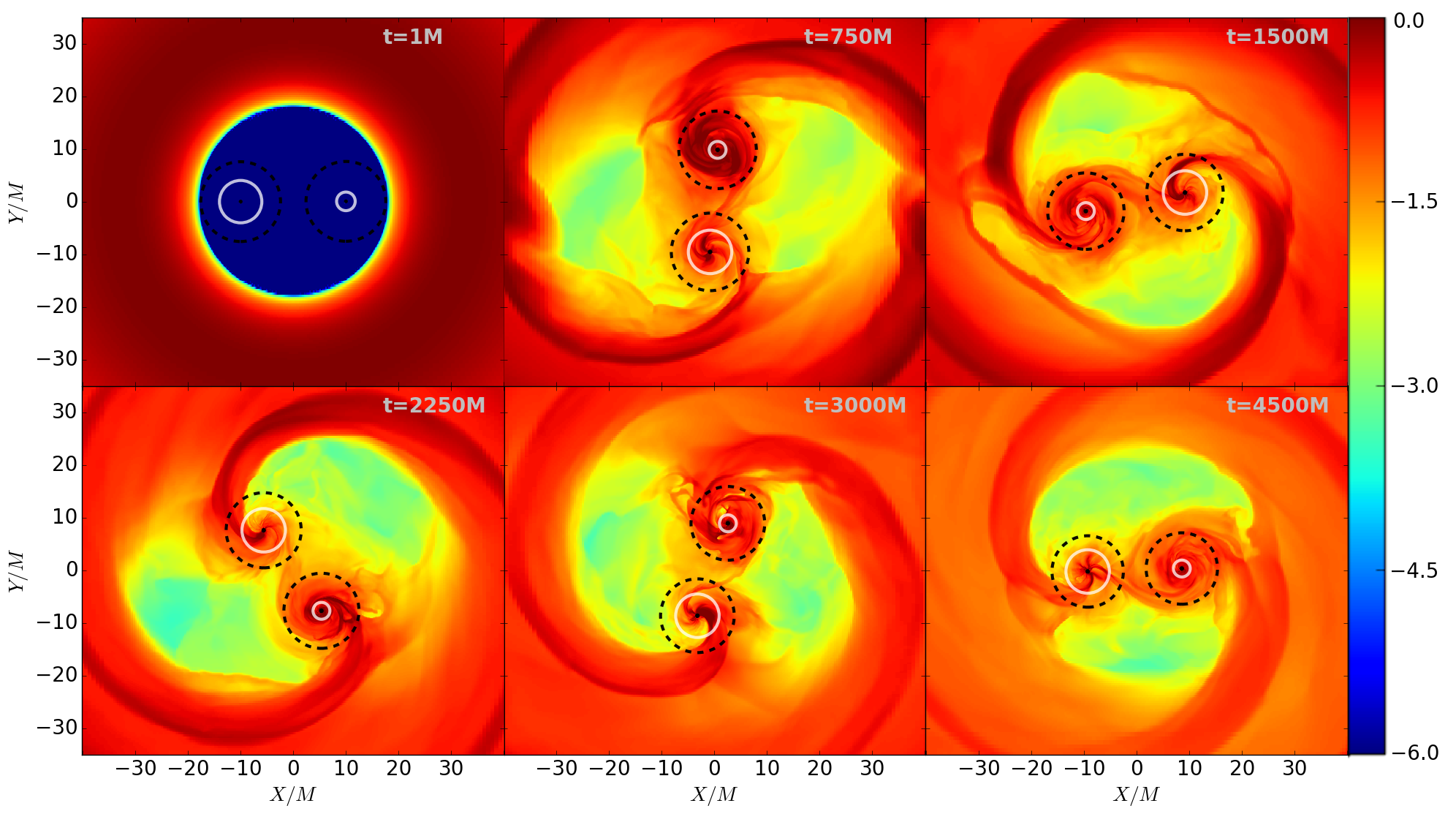}

  \caption{Rest-mass density in the equatorial plane for the
    $\chi_{+-}$ model. A persistent minidisk quickly forms around the
    $\chi_1=+0.75$ black hole, but no disk forms around the
    $\chi_2=-0.75$ black hole. The Hill spheres (black dashed circles)
    and the ISCO radii (white circles) are shown around each black
    hole (assuming each BH was in isolation). For the $\chi_1=+0.75$
    black hole the Hill sphere is significantly larger than the ISCO,
    but for the $\chi_1=-0.75$ they are more comparable in size. 
    \label{fig:updown}}

\end{figure*}

We use the rest-mass density profiles, the mass contained within the
Hill sphere of each black hole, and the accretion rates onto the black
holes (as defined in ~\cite{Farris:2010}) as diagnostics for studying
minidisk formation and evolution. The Hill sphere radius is computed
based on the Newtonian formula $r_{\rm Hill}=0.5 (q/3)^{1/3} d$,
within which we integrate the total rest mass. This definition of the
Hill sphere is based on the coordinate radius, it is not gauge
invariant and is used only as a means to approximate the mass
contained in minidisks. However, since the binary is symmetric this
coordinate radius defines a volume around each black hole that is the
same, which roughly coincides with the outer edges of the minidisks
that form in our simulations as we show below. We also investigate the
temperatures of the minidisks as well as the efficiency of converting
accretion power to luminosity in each of our models.

\begin{figure*}
  \centering
    \includegraphics[width=0.99\textwidth]{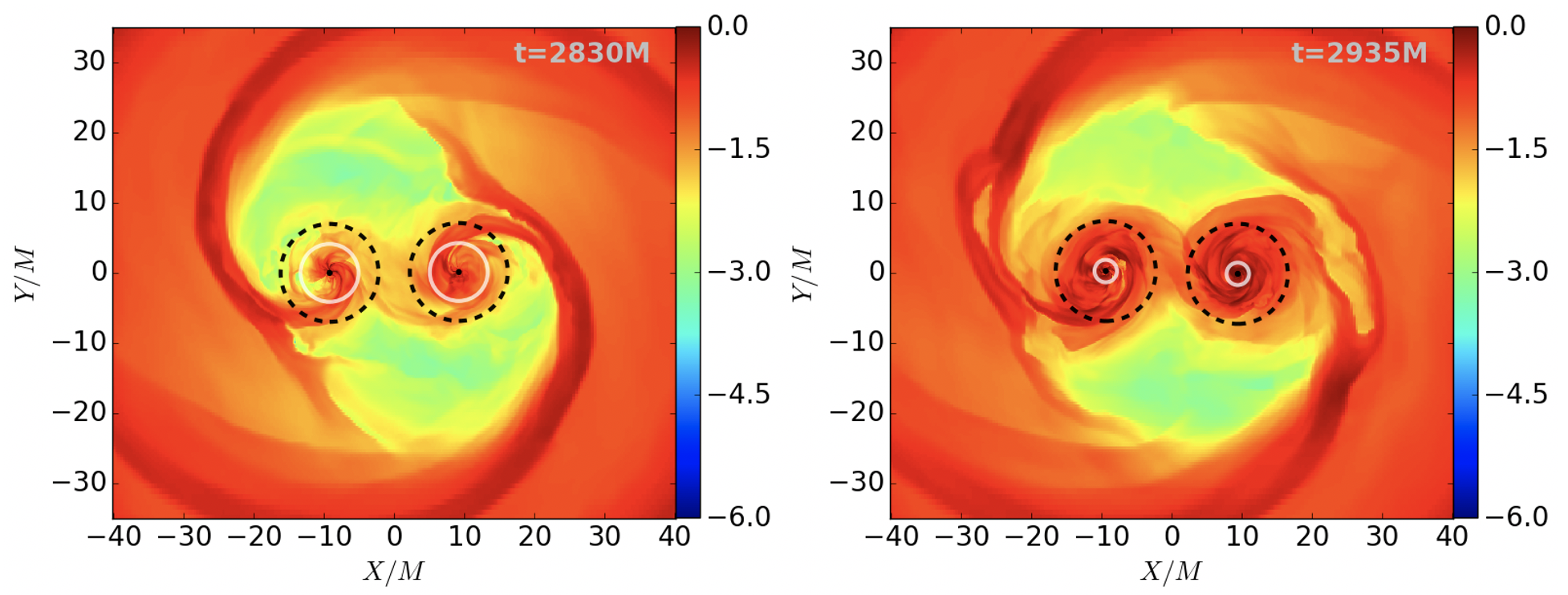}
  \caption{Comparison of the rest-mass density in the equatorial plane
    for $\chi_{--}$ (left panel), and $\chi_{++}$ (right panel), taken
    following completion of five orbits in both cases. Persistent minidisks are seen
    only in the $\chi_{++}$ case as the Hill sphere (black dashed
    circles) is significantly larger than ISCO radius (white circles).
    \label{fig:compare}}
    \end{figure*}

The initial data were designed to probe the interplay of tidal forces
and black hole spins to determine their impact on minidisks. In our
earlier work, we did not find evidence for persistent minidisks.
Minidisks would form and quickly be depleted. In those works the black
holes were nonspinning ($r_{\rm ISCO}=6m=3M$) and the initial
separation was $d=10M$, and hence a Hill sphere of $r_{\rm Hill}\sim
3.5M$, leaving almost no room for stable orbits within the Hill
sphere. By contrast, in this study, the initial separation is $d \sim
20M$, providing a larger Hill sphere of $r_{\rm Hill}\simeq 7M$, which
leaves plenty of room for stable orbits for $\chi=0.75$ black holes,
some room in the $\chi=0$ black holes, and little room in the
$\chi=-0.75$ black holes. Therefore, the expectation is that
persistent minidisks should more easily form dynamically around
$\chi=0.75$ black holes, but not so much around $\chi=-0.75$ black
holes. This is consistent with our findings as we next discuss.

\subsection{Rest-mass Density}
\label{density}




As in our previous studies the matter undergoes a transient phase as
the initial data is relaxed, and the accretion-flow regions closer to
the black holes begin to settle after a time period $\Delta t\simeq
2000M$, as determined by the saturation of the mass accretion rate
onto the black holes $\dot{M}$. This is consistent with typical
findings in single black hole accretion studies (see~\cite{Porth:2019} and
references therein), where it takes a few local orbital periods for
turbulence to fully develop. The accretion flow settles in an
inside-out fashion. Secular trends from an incomplete relaxation at
larger radii should be present, but we do not expect this to affect
our main conclusions. Minidisk structures form even before the
accretion rate saturates and then settle down to a quasi-steady state
around the same time.

In Fig.~\ref{fig:updown} we show equatorial rest-mass density
snapshots from case $\chi_{+-}$. This case is the most unambiguous
demonstration of the effect of spin on minidisk formation. For the
$\chi=+0.75$ black hole $r_{\rm Hill}$ is significantly larger than
$r_{\rm ISCO}$, while for $\chi=-0.75$ black hole $r_{\rm ISCO}$ is
comparable to $r_{\rm Hill}$. As a result, we observe that the
positive spin black hole develops a minidisk while the negative spin
black hole does {\it not}. For the $\chi=-0.75$ black hole, material
that enters the Hill sphere crosses the ISCO (solid white circles)
shortly thereafter, and as a result the spiral streams are accreted
without orbiting the black hole and forming a minidisk. We also
observe that for all cases where persistent minidisks form, the outer
edges of minidisks traces the approximate Hill sphere we draw (dashed
black circles). This set of observations clearly demonstrates the
validity of the argument that minidisks should form only when $r_{\rm
  Hill} \gg r_{\rm ISCO}$, resolving the question under what
conditions minidisks can exist during the late stages of the inspiral.

In addition, we investigate our other models, $\chi_{00}$,
$\chi_{++}$, and $\chi_{--}$ to further support the findings of the
$\chi_{+-}$ model. As expected from our earlier arguments, we find
that cases $\chi_{00}$ and $\chi_{++}$ form persistent minidisks,
while case $\chi_{--}$ ($r_{\rm ISCO}=8.22m=4.11M$) does not, and
exhibits the same behavior as the $\chi=-0.75$ black hole in the
$\chi_{+-}$ model, with spiral streams entering the Hill sphere and
crossing the ISCO, and thereby plunging into the black hole.  In
Fig~\ref{fig:compare}, we show a equatorial rest-mass density snapshot
after five binary orbits have been completed for the $\chi_{++}$, and
$\chi_{--}$, demonstrating the above.

\subsection{Rest mass within Hill spheres and Accretion Rates}
\label{roche_lobes}

\begin{figure*}
\centering 
   \includegraphics[width=0.8\textwidth]{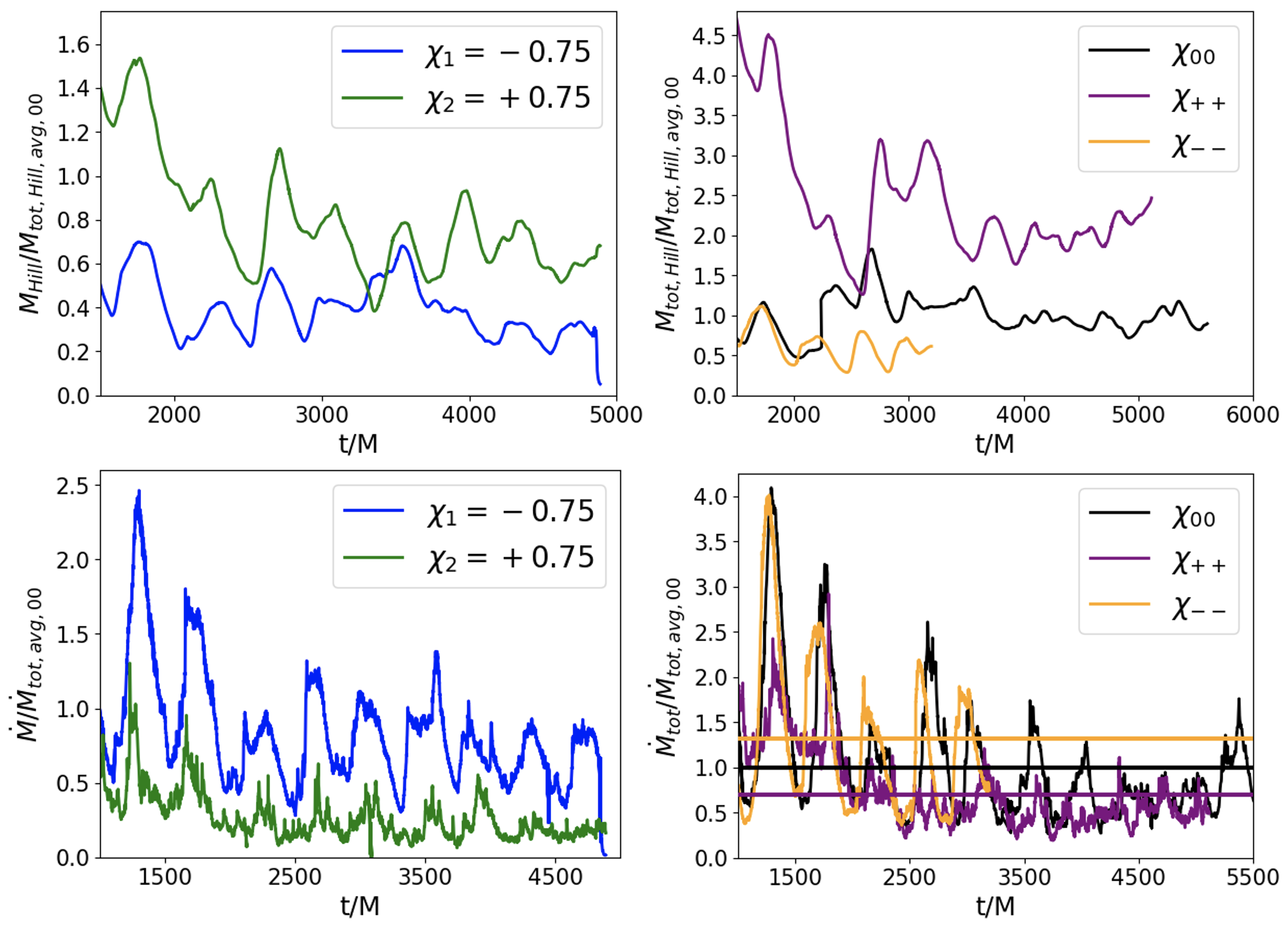}
   \caption{\textit{Top Left:} Rest mass within the Hill sphere of
     each black hole in the $\chi_{+-}$ model as a function of time,
     normalized to the average value fxor the total rest mass in both
     Hill spheres for the $\chi_{00}$ case.  \textit{Top Right:} Total
     rest mass within the Hill spheres of both black holes in each
     model listed as a function of time, with the same normalization
     as in the top left. \textit{Bottom Left:} Rest mass accretion
     rate onto each black hole for the $\chi_{+-}$ model as a function
     of time, normalized to the time-averaged value of the total
     rest-mass accretion rate onto both black holes in the $\chi_{00}$
     case.  \textit{Bottom Right:} Total rest-mass accretion rate onto
     both black holes in each model listed as a function of time with
     the same normalization as in the bottom left. The horizontal
     lines indicate the time-averaged value for each model. $M \approx
     5 (\frac{M}{10^6 M_{\odot}})$ sec.}
   \label{fig:mass+acc}
 \end{figure*}

We next investigate the rest mass contained within the Hill sphere of
each black hole.  The top left panel of Fig~\ref{fig:mass+acc} shows
the rest mass within the individual Hill spheres for the $\chi_{+-}$
case normalized to the time-averaged value of the total rest-mass
within the Hill spheres of the $\chi_{00}$ case. The $\chi=+0.75$
black hole clearly contains more rest mass than the $\chi=-0.75$ one,
demonstrating in an another way that the matter plunges into the
$\chi=-0.75$ black hole as soon as it enters the Hill sphere because
it crosses the ISCO. A similar picture is painted by the top right
panel in the same figure, which shows the total rest mass within the
Hill spheres normalized to the time-averaged value of the total
rest-mass within the Hill spheres of the $\chi_{00}$ case for cases
$\chi_{00}$, $\chi_{++}$ and $\chi_{--}$. As is clear from the image
the total mass within the Hill spheres in the $\chi_{++}$ is about
twice that of the $\chi_{00}$ model, which is about twice that of the
$\chi_{--}$ model.  However, the plots also demonstrate that the
amount of rest-mass within the Hill sphere of even the $\chi=+0.75$
black holes, which form persistent minidisks, oscillates around a mean
value in agreement with the nonspinning simulations reported
in~\citep{Bowen:2017}.

In the bottom left panel of Fig~\ref{fig:mass+acc} we show the
rest-mass accretion rate onto each black hole of the $\chi_{+-}$ model
normalized by the time-averaged total accretion rate of the
$\chi_{00}$ case. The plot clearly shows that the $\chi=-0.75$ black
hole has a higher accretion rate than the $\chi=+0.75$ black hole,
demonstrating the plunging of the accretion streams onto $\chi<0$
black holes, as opposed to orbiting and forming minidisks around
$\chi>0$ black holes. In the bottom right panel we show the total
rest-mass accretion rate normalized by the time-averaged total
accretion rate of the $\chi_{00}$ case for cases $\chi_{00}$,
$\chi_{++}$ and $\chi_{--}$, with the horizontal lines indicating the
time-averaged accretion rate for each case. As expected the
time-averaged rest-mass accretion rate in the $\chi_{--}$ is larger
than that in the $\chi_{00}$ case, which in turn is larger than in the
$\chi_{++}$ case. The $\chi_{00}$, $\chi_{++}$, and $\chi_{--}$ cases
reinforce our findings from the $\chi_{+-}$ model. The accretion rates
exhibit clear periodicities, which will be the focus of a future paper
of ours.

All these findings are consistent with the expectation that the
location of the ISCO has a significant impact on the systems ability
to maintain mass within the Hill spheres and form minidisks toward the
late stages of the inspiral.

\subsection{Jet Luminosities and Temperatures} 

\begin{figure*}[t]
  \centering

  \includegraphics[width=0.99\textwidth]{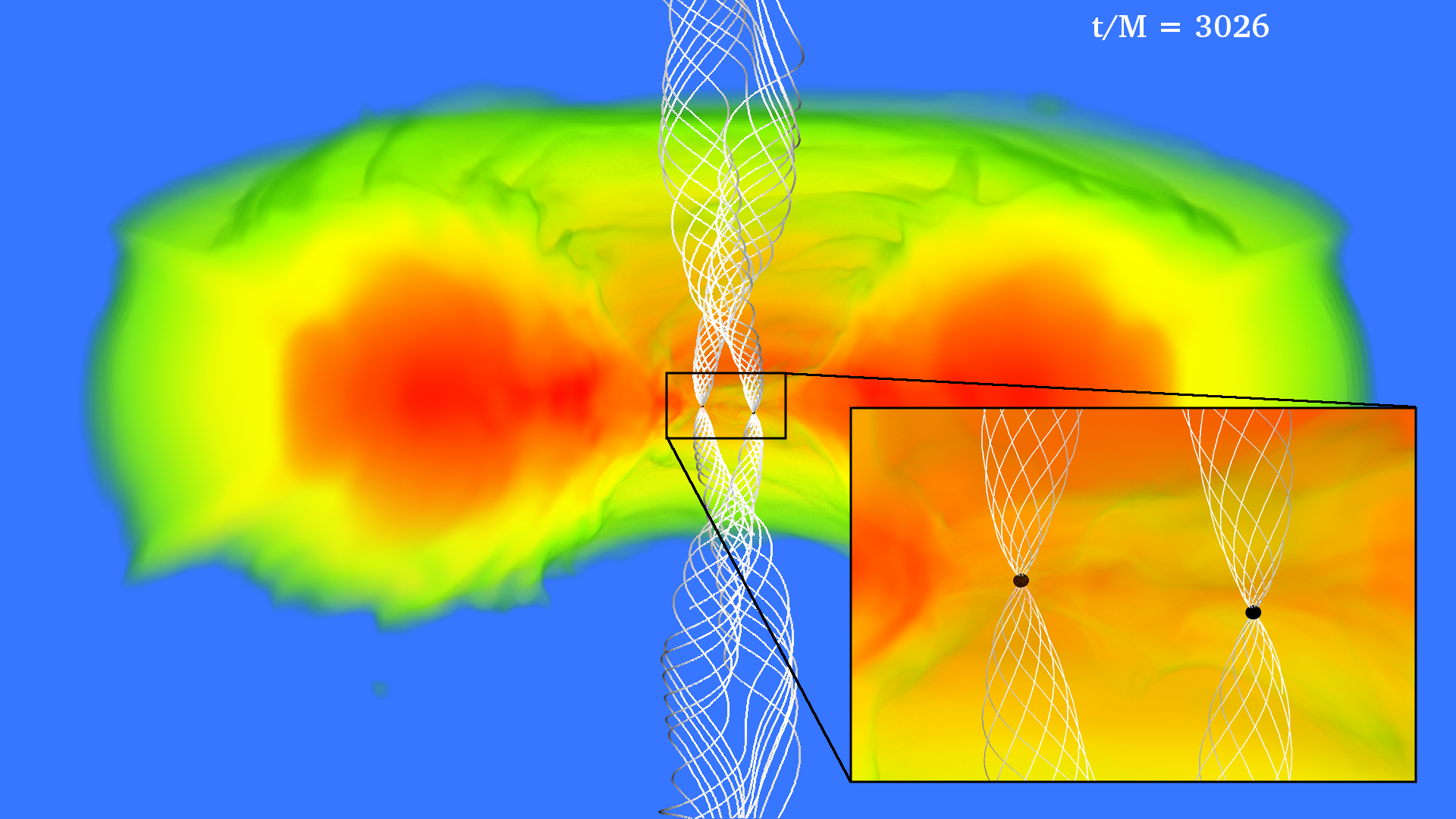}
  \caption{3D view showing rest-mass density (color coded) for the
    $\chi_{++}$ model and magnetic field lines (white) anchored on the
    black hole horizons.  Twin jets are visible above and below each
    black hole (see inset for a zoomed-in view). The black spheres
    indicate the resolved black hole apparent horizons in our
    simulations.
    \label{fig:3d}}

\end{figure*}

We find that jets are launched from both spinning and nonspinning
systems as illustrated in Fig.~\ref{fig:3d}, which shows a 3D
rendering of the rest-mass density of the $\chi_{++}$ model with white
lines indicating the magnetic field lines anchored to the black
holes. The magnetic field lines are more twisted than in the
nonspinning cases we reported in earlier
work~\citep{Gold:2014a,Gold:2014b,Khan:2018ejm} -- a result of black
hole spin. This combined effect leads to a dual jet structure close to
the black holes that merge to form a single jet structure at larger
height.

\begin{figure}[t!]
  \centering \includegraphics[width=0.47\textwidth]{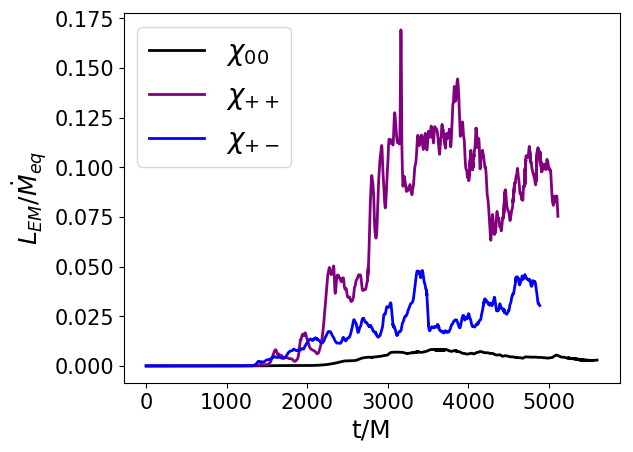}
  \caption{The efficiency for converting accretion power to EM
    luminosity $L_{\rm EM}/\dot{M}_{\rm eq}$ as a function of time. Here
    $L_{\rm EM}$ is the Poynting luminosity and $\dot{M}_{\rm eq}$ is the
    average total rest-mass accretion rate for each model after
    $t=1500M$ when the accretion rate has settled, and $M \approx 5
    (\frac{M}{10^6 M_{\odot}})$ sec.
    \label{fig:em_lum}}

\end{figure}

We calculate the Poynting luminosity associated with the collimated
jet outflow on the surface of coordinate spheres $S$ as $L_{\rm EM} =
\oint_S {T_{0}}^r_{,(\rm EM)}dS$, where ${T_{\mu}}^\nu_{,(\rm EM)}$ is
the EM stress-energy tensor. $L_{\rm EM}/\dot{M}_{\rm eq}$ is the
efficiency for converting accretion power to EM jet luminosity, where
$\dot{M}_{\rm eq}$ is the time-averaged accretion rate after the
accretion rate has settled $(t\gtrsim 1500M)$. We plot the efficiency
as a function of time in Fig.~\ref{fig:em_lum} for the $\chi_{00}$,
$\chi_{++}$, and $\chi_{+-}$ models. We note that it takes time for
the outflow to develop and propagate out the EM luminosity extraction
radius of $150M$, which is why although the flow around the black
holes can relax, it takes longer for the EM luminosity to relax. The
evolution of the $\chi_{--}$ model was long enough for the accretion
rate to relax, but not long enough for the EM luminosity to do so.  As
a result we do not include this model here. The figure clearly
demonstrates that spin plays a significant role in the efficiency of
the luminosity output, with the greatest efficiency achieved in the
$\chi_{++}$ model and the lowest efficiency in the $\chi_{00}$
model. This is the expected outcome resulting from the combination of
the Blandford-Znajek (BZ) effect, which describes outgoing luminosity
produced from a single spinning black hole~\citep{Blandford:1977} (see
also \cite{McKinney:2004ka,Hawley:2006} for GRMHD studies on the spin
dependence of jet luminosities) and an ``orbital'' BZ
effect~\citep{Palenzuela:2010}.

Finally, we compute the effective disk temperature based on the
assumption of radiation pressure dominance (consistent with the the
$\Gamma=4/3$ equation of state), by solving $\rho_0 \epsilon = a
T^4$. We scale the total binary mass to $10^6 M_{\odot}$ and scale the
maximum rest-mass density such that the average accretion rate equals
the Eddington accretion rate with efficiency $10\%$. Doing this
provides a characteristic temperature in the minidisks of around
$T\sim 10^6 \left( \frac{\dot{M}_{avg}}{\dot{M}_{\rm Edd}} \right) ^{1/4}
\left( \frac{M}{10^6 M_{\odot}} \right)^{-1/4}$ K. We find this to be
consistent across each of the models. The minidisks have a higher
temperature than the circumbinary disk, which is itself hottest at the
inner edges and cooler in the bulk of the disk.

\section{Conclusions and Discussion}
\label{conclusions}

We have demonstrated that minidisks in binary black holes form if the
Hill sphere or equivalently the tidal truncation radius is
significantly larger than the ISCO radius. We further showed that the
size of the minidisks (on average) traces the approximate Hill sphere
around each black hole, which implies that the sizes of minidisks are
tidally truncated and therefore a simple, linear function of the
binary separation only. This finding could be important if the size of
minidisks could be inferred from observations. One of the challenges
involved is that source size depends on observation frequency, but one
may hope that correlations between the actual minidisk size and its
photosphere can be found in the future.

Our work establishes a common evolutionary sequence that similar
systems are expected to follow: At large enough separations when the
Hill sphere is larger than the ISCO radius, minidisks are expected to
be present. In this phase the accretion rate could be quasi-periodic
with the size of the minidisks depending linearly on the binary
separation. Ultimately GRMHD simulations in full GR at large
separations are necessary to confirm these expectations and the
presence of accretion periodicities on the binary orbital time. Since
$r_{\rm Hill}=0.5 (q/3)^{1/3} d$ depends on the binary separation, as
the binary inspirals it will reach a threshold where $r_{\rm Hill}
\simeq r_{\rm ISCO}$ and persistent minidisks will no longer be able
to exist and will ``evaporate'' because the tidally stripped accretion
streams from the circumbinary disk are accreted on a short dynamical
time scale as they plunge into the black hole following ISCO
crossing. This leads to more complicated variability in the mass
accretion rate until merger and any EM signature associated with
persistent minidisks is expected to become fainter. Note that the
onset of this transition depends on black hole spin (through the ISCO
radius). We term this anticipated dimming of EM emission from
minidisks ``minidisk evaporation'' and expect it to be an inevitable
outcome, that is qualitatively robust even for different magnetic
field strengths, topologies or initial torus parameters (quantitative
measures such as mini disk accretion rates may well depend on the
details). Following minidisk evaporation toward the late stages of the
inspiral, the qualitative evolution then proceeds in accord with our
previously obtained results
\citep{Farris:2012,Gold:2014a,Gold:2014b,Khan:2018ejm}. Our
simulations also suggest that black hole binaries with prograde spin
maintain minidisks for a longer timescale than nonspinning and
retrograde spin black holes.

These findings could in principle serve as a new diagnostic to probe
black hole spins observationally when combined with information from
the gravitational wave signal. In particular, when LISA is operating
or in the event that Pulsar Timing Arrays detect an individual
supermassive black hole binary, the merger time can be predicted from
the gravitational wave signal. Therefore, the difference of merger
time to the time when the minidisk signature fades away should open a
new avenue to probe black hole spins observationally. For this
strategy to work out additional source modeling and better predictions
from theory will be invaluable. Our work here has shown that when
$r_{\rm Hill} \gg r_{\rm ISCO}$, minidisks can form, but additional
studies to probe the epoch of minidisk evaporation where $r_{\rm Hill}
\gtrsim r_{\rm ISCO}$ will be important to make this a viable and
useful diagnostic.

In addition to the minidisk dynamics here we also found that
jets arising from circumbinary accretion onto binary black holes
toward the late stages of the inspiral are significantly more powerful
when spinning black holes (even with moderately high dimensionless
spin of $\chi=0.75$) are involved. With proper theoretical modeling
this finding also paves a new way to probe black hole spin from future
EM jet observations of these systems.

Finally, apart from the observational implications, our results have
important consequences for future relativistic simulations of these
systems. In particular, the chosen initial orbital separation should
respect the criterion for minidisk formation in order to properly
compute accretion rates and electromagnetic signals, and to faithfully
represent the expected structure of the circumbinary system.

\acknowledgments

This work was in part supported by NSF Grant PHY-1912619 to the
University of Arizona, NSF Graduate Research Fellowship Grant
DGE-1746060, and NSF Grant PHY-1662211 and NASA Grant 80NSSC17K0070
to the University of Illinois at Urbana-Champaign. Computational
resources were provided by the Extreme Science and Engineering Discovery
Environment (XSEDE) under grant number TG-PHY190020. XSEDE is supported
by the NSF grant No.\ ACI-1548562. Simulations were performed on
\texttt{Comet}, and \texttt{Stampede2}, which is funded by the NSF
through award ACI-1540931.

\bibliographystyle{hapj}
\bibliography{ref}

\end{document}